\documentstyle[aps,epsfig]{revtex} 
\begin{document} 
\preprint{} 
\draft

\title{ Sub-Planck spots of Schr\"odinger cats and quantum decoherence}

\author{Wojciech Hubert Zurek, \it Theory Division, T-6, MS B288, LANL, 
Los Alamos, NM 87545, USA}

\maketitle

\begin{abstract}

Heisenberg's principle$^1$ states that the product
of uncertainties of position and momentum should be no less than
Planck's constant $\hbar$. This is usually taken to imply that phase space
structures associated with sub-Planck ($\ll \hbar$) scales do not exist, or,
at the very least, that they do not matter. I show that this deeply ingrained
prejudice is false: Non-local ``Schr\"odinger cat" states of quantum systems
confined to phase space volume characterized by `the classical action'
$A \gg \hbar$ develop spotty structure on scales corresponding to sub-Planck
$a =  \hbar^2 / A  \ll \hbar$. Such structures arise especially quickly
in quantum versions of classically chaotic systems (such as gases, modelled
by chaotic scattering of molecules), that are driven into nonlocal
Schr\"odinger cat -- like superpositions by the quantum manifestations of
the exponential sensitivity to perturbations$^2$. Most importantly, 
these sub-Planck scales are physically significant:
$a$ determines sensitivity of a quantum system (or of a quantum environment)
to perturbations. Therefore sub-Planck $a$ controls the effectiveness 
of decoherence and einselection caused by the environment$^{3-8}$.
It may also be relevant in setting limits on sensitivity of Schr\"odinger
cats used as detectors.

\end{abstract}

\vspace{1cm}

One of the characteristic features of classical chaos is the evolution of
the small scale structure in phase space probability distributions:
As a consequence of the exponential sensitivity to initial conditions,
an initially regular ``patch'' in phase space with a characteristic
size $\Delta $ will, after a time $t$, develop structure on the scales:
$$ z \simeq \Delta  \exp(- \Lambda t) \eqno(1)$$
where $\Lambda$ is, in effect, the Lyapunov exponent$^{9,10}$. This is a
consequence of the exponential stretching and of the conservation
of phase space volume in reversible evolutions.

What happens with small scale structure in quantum versions of
classically chaotic systems? We shall investigate this question
using Wigner function $W(x,p)$ -- the closest quantum analogue of
the classical phase space distribution$^{11}$:
$$W(x,p) = {1 \over {2 \pi \hbar}} \int \exp(ipy / \hbar) \rho(x-{y \over 2},
x +{y \over 2}) dy  \ . \eqno(2) $$
Above, $\rho$ is the density operator. When the quantum state is a pure
$|\psi>$, the integrand becomes $\exp(ipy/\hbar)\psi^*(x+y/2)\psi(x-y/2)$.

In a quantum system the smallest spatial structures in the wave function
$\psi (x)$ will be set by the highest phase space frequencies available.
Given a finite energy, $\psi(x)$, and, hence, $W(x,p)$ cannot reach scales 
as small as the exponential squeezing of Eq. (1) would eventually imply. 
Thus, in a quantum system, there must be a scale below which structure should not appear. That some limit must exist was apparent already some time ago:
Berry and Balazs (who, as we shall see below in connection with Eq. (13))
have an enviable record of correctly anticipating various aspects
of quantum chaos) have conjectured$^{12,13}$ that structure saturates
on scales given by Planck constant. If this were the case, $W$ should be
smooth on scales small compared to $2 \pi \hbar.$

I shall show that copious structure appears in the Wigner function $W$
on much smaller sub-Planck scales associated with the action of the order of:
$$ a \ \simeq \ \hbar * {{\hbar} \over A} \ , \eqno(3) $$
and explain its origin. Most importantly, I shall demonstrate that
$a$ has physical consequences: It determines
sensitivity of the system (or of the environment) to decoherence. Above $A$
is the {\it classical} action of the system, given approximately by the product
of the range of effective support of its state in position $L$ and 
momentum $P$:
$$ A \ \simeq \ L * P \ . \eqno(4)$$
The values of $L$ and $P$ are in turn set by the available energy $E$,
and by the form of the potential $V(x)$ (i. e., $P \leq \sqrt{2 m E},
\ E - V(L) \geq 0$, etc.) that determine the effective support of the
probability distribution in phase space. We shall eventually see that many
of the calculations can be carried out using the state vector of the system
in the appropriate representation. Nevertheless, intuitive understanding of 
the significance of the sub-Planck scale $a$ is easiest to attain starting 
with a more comprehensive view of the phase space image of quantum states
afforded by the Wigner representation.

It is evident from Fig. 1 that -- at least in chaotic systems -- the structure
on sub-Planck scales appears and saturates quickly. Let us now illustrate its
origin: The smallest scales in phase space arise from interference. Consider,
for instance, a familiar ``Schr\"odinger cat" coherent superposition of two
minimum uncertainty Gaussians$^6$:
$$ W(x,p) \ = \  {{G(x+x_0,p) \ + \ G(x-x_0,p)} \over 2} \
+ \ (\pi \hbar)^{-1}
\exp(-{{p^2\xi^2} \over \hbar^2} - {x^2 \over \xi^2}) \cos (p{{2 x_0 
} \over \hbar}) \ ,
\eqno(5)$$
where Wigner functions of the two Gaussians
`east' and `west' of the center are;
$$ G(x \pm x_0, p-p_0) \ = \ (\pi \hbar)^{-1} {{ \exp({ - (x \pm x_0)^2/\xi^2
- (p-p_0)^2 \xi^2/\hbar^2 }} }) \ . \eqno(6) $$
The last oscillatory term $W_{WE}$ in Eq. (5) is a symptom of interference.
Its ripples have a frequency proportional to the separation $L=2x_0$ between
the two peaks. When we define the frequency of the ripple pattern in
momentum $f_p$ through $ \cos (L p / \hbar) = \cos (f_p ~ p) ,$ then
$ f_p  \ = \ {L / \hbar}  . $
Ridges and valleys of such interference pattern are  always parallel to the
line of sight between the two Gaussian peaks. Thus, standing on top of one
Gaussian peak, one could still see the other peak through the valleys (and
between the ridges) of the interference term, even though its envelope is
a factor of two higher than either Gaussian.

Coherent states form an overcomplete set.
We can therefore express an arbitrary pure state as a superposition of
coherent states.
The smallest interference structures in such an expansion arise from pairs
of coherent states separated by the whole range available to the system in
phase space. Therefore, we expect smallest scales with:
$$ f_x = P/\hbar, \ \ {\rm or} \ \delta_x  \ = \ \hbar / P \ , \ \ \ \
f_p = L/\hbar, \ \ {\rm or} \  \delta_p \ = \ \hbar / L \ . \eqno(7,8)$$

As an example, consider {\it compass state}, a Schr\"odinger cat-like
superposition of four minimum uncertainty Gaussians, one pair located north
and south of the common center, the other east and west (see Fig. 2). Wigner
distribution is quadratic in the wave function. Therefore, $W$ of any
superposition can be reconstructed from the contributions corresponding to, at
most, pairs of states. The structure of Wigner distribution of a superposition
of a pair of Gaussians, (Eq. (5) that is, {\it nota bene}, reflected in
the patterns on the sides of the square in Fig. 2), can be then usedto infer $W$ of the compass state:
$$W_{NWSE} \ = \ (G_N \ + \ G_W \ + \ G_S \ + \ G_E)/4  \eqno(9a)$$
$$ +  \ (W_{NW} \ + \ W_{WS} \ + \ W_{SE} \ +\ W_{EN})/2 \eqno(9b)$$
$$ + \ (W_{NS} \ + \ W_{EW})/2 \  \eqno(9c)$$
in an obvious ``geographic'' notation. The last line of Eq. (9);
$$ W_{NS} + W_{EW} \ = \ (\pi \hbar)^{-1} \exp(-{{p^2 \xi^2} \over \hbar^2}
- {x^2 \over \xi^2} )
\bigl(\cos{{pL} \over \hbar} \ + \ \cos{{xP} \over \hbar} \bigr) \ . 
\eqno(10)$$
is of greatest interest.
Above, we have assumed that shapes of all the Gaussians
are identical, so that the exponential envelope in Eq. (9c) is common
to both terms.

The resulting interference term (the center of Fig. 2) has a
``checkerboard pattern''. The size of the single `tile' can be obtained
from zeros of the oscillatory factor of Eq. (10);
$$\cos(pL/\hbar) \ + \ \cos(xP/\hbar) \ =  \ 2\cos{{Px + Lp} \over {2 \hbar}}
\cos{{Px - Lp} \over { 2\hbar}} \ . \eqno(11) $$
Individual squares, four per tile, are defined by zeros that occur when:$ x = \pm \pi \hbar /  2L , \ p = \pm \pi \hbar / 2 P .$
The fundamental periodic tile has an area of:
$$ a = {{ 2 \pi \hbar} \over L} * {{ 2 \pi \hbar} \over P } = (2 \pi \hbar)^2/A
  \ ,  \eqno(12)$$
that, with the identification of $A = L * P$, yields
action associated with the smallest scales present in quantum phase space.

The above calculation shows by construction that a quantum state spread over
a phase space of volume $A=L*P$ can accommodate Wigner distribution structures
as small as $a$. Such states arise naturally: Evolution will
force almost any system (with a notable exception of a harmonic oscillator)
into a Schr\"odinger cat state$^2$ -- a coherent non-local
superposition that, after a time it takes the wave function to spread over
the phase space volume $A$ -- inevitably develops interference pattern
on the scale given by Eqs. (7-10).

It is not necessary to invoke chaos: After sufficient time, even
an integrable non-linear system may spread coherently throughout the
available phase space, and, consequently, saturate small scale structure.
It is just that when the evolution is chaotic, such small scales
will be attained faster, on a time scale given by$^{14}$:
$$ t_{\hbar} \ = \ \Lambda^{-1} \ln \Delta p \chi / \hbar \ . \eqno(13a) $$
Above, $\Delta p$ is the characteristic spread of the initial smooth
probability distribution. $\chi$ characterizes the scale on which potential
is significantly nonlinear. It is typically given by
$\chi \simeq \sqrt {V' / V'''}$. Similar estimates obtain from the formula:
$$ t_r \ = \ \Lambda^{-1} \ln A / \hbar \eqno(13b)$$
deduced some time ago$^{12,15}$.

There is one more suggestive way to express saturation scale $a$: The number
of distinct (orthogonal) states that can fit within phase space of volume $A$
is ${\cal N} = A/(2 \pi \hbar)$. The structure  we are discussing appears
therefore on the scale $a \sim {{\hbar}/{\cal N}} .$
Here ${\cal N}$ is, in effect, the dimension of the available Hilbert space.

In accord with Heisenberg's principle, quantum system cannot be localized
to a sub-Planck volume in phase space. Hence, one might be tempted to dismiss
sub-Planck scales as unphysical even if they appear in the Wigner distribution.
In particular, if a state cannot be confined, by measurements, to a volume
less than $\sim \hbar$, then one might expect that it will not to be noticeably
perturbed by displacements much smaller than $\sqrt \hbar$. We now show
that this expectation is false, and that $a \ll \hbar$ plays a decisive role
in determining sensitivity of quantum systems (or of quantum environments) to
perturbations: Phase space displacements $\delta \sim \sqrt a \ll \sqrt \hbar$
shift the state with a dominant structure on scale $a$ enough to make it
orthogonal -- i.e., distinguishable -- from the unshifted original.
Sensitivity to perturbations in turn sets the limit on the efficiency
of decoherence.

There are two complementary aspects to this connection between $a$ and
decoherence: When the system with a sub-Planck scale in $W$ is coupled to the
environment, decoherence can be thought of as monitoring, by the environment,
of some of its observables$^{3-8}$. Its effect -- suppression of quantum
coherence -- can be traced to the Heisenberg's principle: The observable
complementary to the one monitored by the environment become less determined,
in effect smearing Wigner distribution along the corresponding
phase space direction$^{8}$. When this smearing obliterates interference
structures on scale $a$, coherence on the large scales corresponding to
$A \sim \hbar^2/a$ shall be suppressed, and the Schr\"odinger cat state
would have lost its quantum nonlocality$^{2,14,17}$.

A complementary aspect of the same story will be our focus in the remainder
of this paper. It involves the situation when the state of the environment
-- the cause of decoherence -- is a ``Schr\"odinger cat" spread over
the phase space region $A$. Decoherence is caused by `monitoring' of the
system by the environment. Environment entangles with the system, acting
as an apparatus$^{3-8}$. The sensitivity of the environment to 
perturbations is therefore of essence. We shall test sensitivity
of such Schr\"odinger cat environment by allowing it to interact with a
Schr\"odinger cat system, assumed to be initially in a superposition of
two perfect pointer states $\{ |+>,|-> \}$. Pointer states, taken one at 
a time, perturb the state of the environment but do not entangle with it. 
However, each pointer state perturbs the environment differently$^3$. 
Therefore, a system prepared in a general superposition state will 
leave a pointer state dependent imprint on the environment, and, hence,
entangle with it. We shall show that the displacement $\delta \sim \sqrt a$
sets the size of the smallest perturbations distinguished by the environment,
which in turn controls its ability to decohere the system. Thus, when $|+>$ 
and $|->$ shift the state of the environment differently,
the off-diagonal terms of the density matrix of the system will be
suppressed by a factor $|<\varepsilon_+|\varepsilon_->|$, where:
$$|<\varepsilon_+|\varepsilon_->|^2 =  2 \pi \hbar \int W_+ W_- dx dp
\eqno(14)$$
Above, kets $|\varepsilon_{\pm}>$ and their Wigner distributions
($W_{\pm}$) represent states that evolve from
the original state of the environment $|\varepsilon>$ due to the interaction
(induced, say, by the system-environment interaction Hamiltonian such as
$ \sim (|+><+| ~ - ~ |-><-|) {\hbar \over i} \partial_x$
with the states $|\pm>$ of the system, respectively.

Let us now demonstrate how the behavior of the magnitude of the overlap;
$$|<\varepsilon_+|\varepsilon_->| = |\int \varepsilon^*(x)
e^{i\delta_px/\hbar} \varepsilon(x + \delta_x) dx | \ ,  \eqno(15)$$
is controlled by $a$. Above, $\vec \delta = (\delta_x, \delta_p$) is the
net displacement between $|\varepsilon_+>$ and $|\varepsilon_->$
corresponding to $W_+=W(x+\delta^+_x, p+\delta^+_p)$ and
$W_-=W(x+\delta^-_x, p+\delta^-_p)$, respectively. The displacement
$\vec \delta$ is the difference between the shifts caused by $|+>$ and $|->$,
$\vec \delta = \vec \delta^+ - \vec \delta ^- \neq 0$.
Heuristic argument for the size of displacement that causes orthogonality
$<\varepsilon_+|\varepsilon_-> \approx 0$,
and, hence, decoherence, is easiest to follow when phrased in terms of Wigner 
functions: Suppose $W_+$ and $W_-$ in Eq. (14) have a small scale structure with
patches of alternating sign, as seen in Figs. 1-3. Integral of their product
can reach a maximum value of unity only when $W_+$ and $W_-$ are not shifted
with respect to each other. For shifts small compared to the typical size
of the patches the integrand will still be almost everywhere positive,
but with the increase of the magnitude of the shift the integral will
decrease, as $W_+(x,p) W_-(x,p)$ is no longer
positive definite. As the shift exceeds the size of the patches in $W(x,p)$,
the integrand will oscillate around zero, and the integral of Eq. (14) will
be small compared to unity. If the interference pattern in the Wigner
distribution is periodic (as is the case in Fig. 2) the oscillation will be
also periodic with a period related to the size of the fundamental 
`tile'. When, however, patches are random (as for the typical case 
illustrated in Fig. 3), the
overlap -- having decayed after a displacement $|\vec \delta | \sim \sqrt a$
-- shall not significantly recur. A more formal version of this heuristic
argument is put forward and backed up by numerical simulations elsewhere
(Karkuszewski et al, in preparation). A simple back-of-the-envelope calculation
valid for a generalization of compass states (that is, when the state of
the environment can be approximated by a ``sparse" collection of identically
shaped minimum uncertainty Gaussians) is given in the caption of Fig. 3.

This intuitive picture, already supported the example of Fig. 2, can be further
confirmed by a general yet straightforward calculation based on Eq. (15).
To simplify notation we consider the case when $\vec \delta = (0, \delta_p)$,
i.e., when the net shift is aligned with one of the axes.
(This assumption can be made with no loss of generality, as the axes in phasespace can be rotated to align one of them with an arbitrary
$\vec \delta$.) In that case:
$$ <\varepsilon_+|\varepsilon_-> = \int |\varepsilon(x)|^2 e^{i {{\delta_p x}
\over  \hbar}} dx \ .\eqno(16)$$
This is a simple, general, and compelling result: Suppression the off-diagonal
terms in the density matrix of the system is given by a Fourier transform
of the {\it probability distribution} in the environment along the direction
perpendicular to the net relative shift induced by its coupling with 
the system. This leads back to sub-Planck scales:
For small displacements the exponent in the integrand can be expanded. This
yields:
$$ |<\varepsilon_+|\varepsilon_->|^2 \approx 1 -
\delta_p^2(<x^2>-<x>^2>)/\hbar^2  \ . \eqno(17a)$$
The spatial extent of the wave function is naturally defined as
$L = \sqrt{<x^2>-<x>^2}$,
where the averages of the observable $x$ are over the state of the environment.
The estimate of the shift $\delta_p$ leading to orthogonality is then;
$$ \delta_p \simeq \hbar / \sqrt{<x^2>-<x>^2} \ , \eqno(17b)$$
in accord with the estimate of the dimensions of the sub-Planck structures,
Eq. (8), and in agreement with our thesis about their role. I show in
{\it Methods} that the same simple formula holds when the
environment is initially in a mixed state.

Obvious generalizations of Eqs. (16) and (17) are valid for arbitrary
displacements, and for the case of many states of the system. Our treatment
can be also extended to interactions between the system and the
environment that cannot be represented as simple shifts, although 
calculations become more complicated.
The role played by the spread $L$ of the state of the environment in the
behavior of the overlap (and, hence, in decoherence) is a direct consequence
of the properties of the Fourier transform. Detailed analysis of other
consequences of Eq. (16) is beyond the scope of this paper. Using elementary
properties of the Fourier transform readers can nevertheless 
immediately confirm
that the overlap decreases to near zero around $\delta_p$, Eqs. (8) and (17),
and that -- for typical initial states of the environment -- it remains much
less than unity. ``Revivals" of the overlap we have seen in the case of the
compass state (Fig. 2) are now easily understood and dismissed as an exception.
Equation (16) shows that they can happen only when the initial probability
distribution of the environment in the direction perpendicular to displacement
is localized to a few peaks, so that $|\varepsilon(x)|^2$ --  the spectrum
of the displacement-dependent overlap -- is essentially discrete.

To conclude, a physical consequence -- orthogonality -- can be induced by
displacements much smaller than these corresponding to the
Planck scale $\sqrt \hbar$. This is surprising, as sub-Planck scales are often
regarded as unphysical. We have shown that they have dramatic
physical implications: Displacements given by the square root of $a$, Eq. (3);
$ \delta \simeq \hbar / \sqrt A  , $ suffice to induce orthogonality.
They are a factor $\sim \sqrt{\cal N} = \sqrt{A / 2 \pi \hbar}$ smaller
than $ \delta \sim \sqrt {\hbar}$ dictated by the Planck constant and needed
to move a typical minimum uncertainty Gaussian in a random direction enough
to noticeably reduce its overlap with its (old) self.

Nonlinear, and, especially, chaotic dynamics leads to states that are spread
over much of the available phase space$^{2,12-15}$. Hence, one would expect
that environments with unstable dynamics will be much more efficient 
decoherers,
as they are constantly evolving into delocalized states. Yet, standard
models of decoherence$^{18-20}$ employ harmonic oscillators. They make up for
this inefficiency by using many (infinity) of them, so that
each becomes slightly displaced by the interaction with the system.

On a more mundane level, our results allow one to anticipate 
structure of the mesh required to simulate
evolution of a quantum system in phase space. They are related to
the quantization of discrete chaotic maps on a torus$^{21}$, which turn out to
require a mesh similar to this given by Eqs. (7) and (8). This is in contrast
to simulations of classical systems, that are doomed by Eq. (1), which implies
resolution exponentially increasing with time.
Most importantly, $a$ controls sensitivity of quantum states to perturbations,
with obvious implications for decoherence we have outlined above. Moreover,
sensitivity to perturbations will set limits on `hypersensitivity' of quantum
chaotic systems$^{22}$, and helps one understand enhanced
capacity of quantum chaotic systems for entanglement$^{23}$.

It is tempting to imagine that the sensitivity of quantum systems in highly
delocalized states may be not just a cause of accelerated decoherence (and, 
hence, an impediment to truly quantum applications) but that in certain settings
it may be beneficial. This is not be as far-fetched as it may seem at first:
After all, a detector in a compass-like state of Fig. 2 would be
sensitive to perturbations $\sim \hbar / \sqrt A$ that are minute compared
with the `standard quantum limit'$^{24}$. And, when the to-be-detected
weak force perturbs its state enough to make it orthogonal to the initial
state, an in principle distinguishable record has been made. Sensitivity of
quantum meters would be then limited by $a$, in accord with Eqs. (3), (16),
and (17). Thus, it seems possible that the Schr\"odinger cat may eventually
follow the by now beaten path of other paradoxical quantum 
{\it gedankenexperiments}
that find -- after dusting off of the classical preconceptions and shedding
the aura of the paradox -- a far more useful employ in potential applications
of quantum physics.

\bigskip

\centerline{\bf Methods}

Decoherence happens to a quantum system ${\cal S}$ as a consequence of a
measurement-like
interaction with the environment ${\cal E}$, which entangles their states:
$$ |s>|\varepsilon > = (\alpha |+> + \beta |->) |\varepsilon> \longrightarrow
\alpha |+> |\varepsilon_+ > + \beta |->) |\varepsilon_->
= |\Phi_{\cal SE}>\ .$$
Above, we have assumed a system with a two-dimensional Hilbert
space spanned by the  $\{ |+>,|-> \}$ orthonormal basis. The two conditional
states of the environment $|\varepsilon_{\pm}> = U_{\pm} |\varepsilon>$
evolve under the unitary transformations $U_+ ~( U_- )$ induced by the system
in the state $|+> (|->)$ respectively. When the effect of the interaction is
simply a displacement in phase space, then
$U_{\pm}=D_{\pm} = \exp i(\delta^{\pm}_x p + \delta^{\pm}_p x)/\hbar$,
where $D$ is the displacement operator, and $\vec \delta^{\pm}=(\delta_x^{\pm},
\delta_p^{\pm})$ are the resulting shifts. Such displacement could be 
induced by a Hamiltonian of interaction $H_{\cal SE} = g (|+><+|-|-><-|)
{\hbar \over i} \partial_x$, where $g$ is a coupling constant.
More general conditional evolutions
of ${\cal E}$ can be of course considered.

Following entanglement, the state of the system alone is described by
the reduced density matrix obtained from  $|\Phi_{\cal SE}>$ by
a trace over the environment:
$$ \rho_{\cal S} = Tr_{\cal E} |\Phi_{\cal SE}> <\Phi_{\cal SE}| = $$$$|\alpha|^2 |+><+| \ +  z \alpha \beta^* |+><-| \ +
z^* \alpha \beta^* |-><+| \ + |\beta|^2 |-><-| $$
Disappearance of the off-diagonal terms signifies perfect decoherence.
In the  $\{ |+>,|-> \}$ basis this is guaranteed when the overlap
$z= <\varepsilon_+|\varepsilon_-> = Tr|\varepsilon_-><\varepsilon_+|$
disappears. It is therefore natural to measure effectiveness of decoherence by
the magnitude of the overlap of the two conditional states of the environment,
that in turn determines the degree of suppression of the off-diagonal terms
in $\rho_{\cal S}$.

This two paragraph ``crash course" is no substitute for a more complete
discussion of decoherence$^{6-8}$. We have swept under the rug a number of
issues. Foremost among them is {\it einselection} -- the emergence, in course
of decoherence, of the preferred set of {\it pointer states} that habitually
appear on the diagonal of $\rho_{\cal S}$ essentially independently of
the initial states of either ${\cal E}$  or ${\cal S}$. Stability of
these pointer states (rather than the diagonality of the density matrix
in some basis) is the key to the role played by decoherence in transition
from quantum to classical$^{2-8,19}$.

Another important subject we have avoided in the paper is the likely situation
when the state of the environment is represented by a mixture $\rho_{\cal E}$.
Detailed discussion of this case (treated extensively before$^{6-8,18-20}$,
although not from the point of view of the sub-Planck structures)
in the present context is beyond the scope of this letter, but the basic
conclusion is easy to state: The estimated magnitude of the smallest
displacement leading to orthogonality is still given by Eq. (17).
That is, it is still related to the smallest scales compatible with the
classical action $A$ associated with $\rho_{\cal E}$. Off-diagonal terms of
$\rho_{\cal S}$ are suppressed by $z= Tr U_- \rho_{\cal E} U_+$, which is
the relevant generalization of Eq. (16). This expression can be expanded for
simple displacements $U_{\pm}=D_{\pm}$ in the limit of small shifts to recover
Eq. (17), with the only difference arising from the fact that now the mixture
$\rho_{\cal E}$ must be used to obtain the averages
(e. g. $<x^2>=Tr x^2 \rho_{\cal E}$, etc.).

Note that throughout the paper we have set our discussion
in one spatial dimension. Generalization to
$d$ dimensions is as straightforward conceptually as it is notationally
cumbersome. The structure saturates in volumes of $a^d$, etc. This has little
effect on decoherence, as it depends on displacements that yield
orthogonality, and these are still $\delta \sim \sqrt a $.

\bigskip
\noindent{\bf References}

[1] Heisenberg, W., \"Uber den anschaulichen Inhalt der quantentheoretischen
Kinematik and Mechanik, {\it Z. Phys.} {\bf 43}, 172-198 (1927). English
translation `The physical content of quantum kinematics and mechanics' in
Wheeler, J. A., and Zurek, W. H., {\it Quantum Theory and Measurement}
(Princeton University Press, Princeton, 1983).

[2] Zurek, W. H., Decoherence, chaos, quantum-classical correspondence, and
the algorithmic arrow of time, {\it Physica Scripta} {\bf T76} 186-198 (1998).

[3] Zurek, W. H., Pointer basis of a quantum apparatus: Into what mixture
does the wavepacket collapse? {\it Phys. Rev.} {\bf D 24}, 1516-1524 (1981);

[4] Zurek, W. H., Environment-induced superselection rules,
{\it Phys. Rev.} {\bf D 26}, 1862-1880 (1982).

[5] Joos, E., and Zeh, H. D., The emergence of classical properties through
the interaction with the environment, {\it Zeits. Phys.} {\bf B 59}, 
229 (1985).

[6] Zurek, W. H., Decoherence and the transition from quantum to classical
{\it Physics Today} {\bf 44}, 36-46 (1991).

[7] Giulini, D., Joos, E., Kiefer, C., Kupsch, J., Stamatescu, I.-O., and
Zeh, H. D., {\it Decoherence and the Appearance of a Classical World in Quantum
Theory}, (Springer, Berlin, 1996).

[8] Zurek, W. H., Decoherence, einselection, and the quantum origin of
the classical, {\it Rev. Mod. Phys.}, submitted (2000) ({\tt quant-ph 010527}).

[9] Haake, F., {\it Quantum Signatures of Chaos} (Springer, Berlin, 1991)

[10] Casati, G., and Chrikov, B., {\it Quantum Chaos} (Cambridge University
Press, Cambridge, 1995).

[11] Hillery, M., O'Connell, R. F., Scully, M. O., and Wigner, E. P.,
Distribution functions in physics: Fundamentals, {\it Phys. Rep.} {\bf 106},
121-167 (1984).

[12] Berry, M. V., and Balazs N. L., Evolution of semiclassical quantum states
in phase space, {\it J. Phys. A} {\bf 12}, 625-642 (1979).

[13] Korsch, H. J., and Berry, M. V., Evolution of Wigner's phase-space density
under a nonintegrable quantum map, {\it Physica} {\bf D3} 627-636 (1981).

[14] Zurek, W. H., and Paz, J. P., Decoherence, chaos, and the Second Law,
{\it Phys. Rev. Lett.} {\bf 72}, 2508-2511 (1994).

[15] Berman, G. P., and Zaslavsky, G. M., {\it Physica} (Amsterdam) {\bf 91A},
450 (1978).

[16] Karkuszewski, Z., Zakrzewski, J., and Zurek, W. H., Breakdown of
correspondence in chaotic systems: Ehrenfest versus localization times, e-print
{\tt quant-ph/0010011}


time-dependent


[17] Habib, S., Shizume, K., and Zurek, W. H., Decoherence, chaos, and the
correspondence principle {\it Phys. Rev. Lett}, {\bf 80}, 4361 (1998).


Entropy signatures

[18] Caldeira, A. O., and Leggett, A. J., {\it Physica} {\bf 121A}, 587-616
(1983).

[19] Paz, J. P., and Zurek, W. H., Environment-induced decoherence and the
transition from quantum to classical, Les Houches Lectures, in press (2000).

[20] Braun, D., Haake, F., and Strunz, W. A., Universality of decoherence,
{\it Phys. Rev. Lett.} {\bf 86}, 2913-2917 (2001).

[21] Hannay, J. H., and Berry, M. V., Quantization of linear maps on a torus
-- Fresnel diffraction by a periodic grating {\it Physica} {\bf 1D}, 267-290
(1980).

[22] Caves, C. M., Information, entropy, and chaos, pp. 47-77 in
{\it Physical Origins of Time Asymmetry}, edited by J. J. Halliwell,
J. P\'erez-Mercader, and W. H. Zurek (Cambridge University Press, Cambridge,
1993).

[23] Miller, P. A., and Sarkar, S., {\it Phys. Rev.} {\bf E60}, 1542 (1999).

[24] Braginsky, V. B., and Khalili, F. Y., {\it Rev. Mod. Phys.} {\bf 95},
703-711, (1996).


%


\bigskip

\noindent{\bf Acknowledgments} This research was supported in part by NSA.
Useful chaotic conversations with A. Albrecht, N. Balazs, C. Jarzynski, 
Z. Karkuszewski, and J. P. Paz are gratefully acknowledged.

\begin{figure}[h]
\centering
\epsfig{file=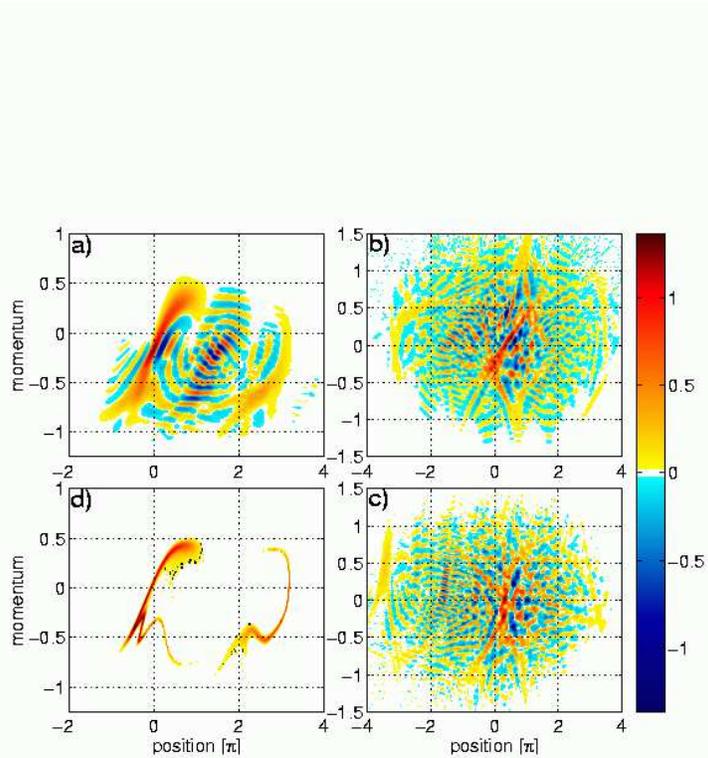, width=10cm, clip=}
\caption{ 
Snapshots of the classical probability density and
of the quantum Wigner distribution in phase space of an evolving chaotic
system with the Hamiltonian:
$H\ = \ {p^2 \over {2 m}} - \kappa \cos (x-l \sin t)+{{a x^2} \over 2} .$
For parameters $m=1, \  \kappa=0.36, \ l=3,$ and $a = 0.01$ this system
exhibits chaos$^{16}$ with Lyapunov exponent $\Lambda \approx 0.2$. Initial
probability density was given by the same Gaussian in both the quantum
(a-c; $\hbar=0.16$) and the classical (d) cases.  Structure on sub-Planck
scales appears and saturates quickly.
Note the contrast between the smallest dimensions of the probability density
in the classical case (d) and of the Wigner distribution at the corresponding
time (a). Exponential shrinking of the smallest scales makes it impossible to
simulate accurately (e.g., reversibly) classical evolution much beyond the
time shown in panel (d) above. By contrast, structures in quantum Wigner
distribution saturate on scales $a \simeq \hbar ^2/A$ soon after $t\sim 20$
(which is of the order of the estimated $t_{\hbar}$, Eq. (13a)). The scale of
structure saturation of a state can be inferred from the volume of the domain
containing it: Smallest structures have dimensions $\delta_x = \hbar/P$
($\delta_p = \hbar / L$), where $P$ and $L$ defines the extent of the envelope
of the effective support of the state ($P^2 \simeq <p^2> - <p>^2$,
$L^2 \simeq <x^2>-<x>^2$). Action associated with the smallest structures will
be then $a \simeq \hbar^2/LP \simeq \hbar^2/A$ in one spatial (two phase space)
dimensions. Generalization to the case of may dimensions is straightforward.
One way to understand structure saturation is through the menu of wavevectors
available in the system that has momenta restricted to range $P \sim \sqrt E$,
where $E$ is the total energy. Complementary argument ($V(L) \sim E$, where
$V(x)$ is the potential) can be made for the smallest scale of ``corrugation"
of the Wigner distribution in $p$.}
\end{figure}

\newpage

\begin{figure}[h]
\centering
\epsfig{file=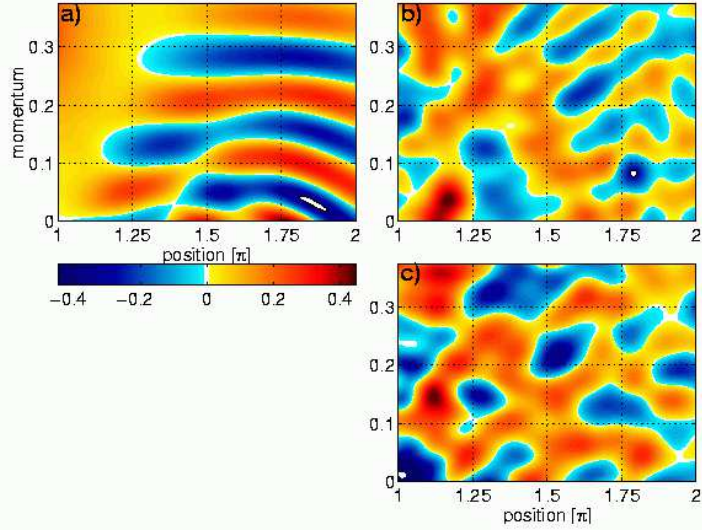, width=10cm, clip=}
\caption{
The compass state, Eq. (11), is a Schr\"odinger cat
- like superposition of four minimum uncertainty states
($|N>,|W>,|S>,|E>$) placed `north', $\dots$ `east', of the common center.
The form of the central interference pattern in $W_{NWSE}$ can be inferred
from the familiar structure of the superposition of two Gaussians, Eq. (5),
also apparent on the sides of the square above: The interference terms
corresponding to the north-south and east-west pairs superpose, creating
a checkerboard pattern. The area of individual tiles is set by the
dimensions of the $NWSE$ cross and corresponds to the classical action $A$, 
of phase space area of the effective support of its envelope, Eq. (12).
Appearance of this interference pattern explains the origin of the structure
saturation seen in Figs. 1 and 3: A system that can be effectively confined
to phase space volume $A$ cannot develop structure on scales smaller
than $a$. Sensitivity of the compass state to perturbations is controlled
by $a$. This is readily seen in the {\it sparse limit}, that is when
when $L\gg \xi, \ P\gg \hbar/\xi$. For simplicity, consider a shift
$\vec \delta = \delta_x \hat x + \delta_p \hat p$ small compared to
the sizes of the Gaussians, $\delta_x \ll \xi$, $\delta_p \ll \hbar/\xi$.
The square of the overlap of the original and displaced states
$|<\psi_{NWSE}|\psi_{NWSE}^{\delta}>|^2 = 2 \pi \hbar
\int W_{NWSE}(x,p) W_{NWSE}(x+\delta_x,p+\delta_p) dx dp $ is approximately
$|<N|N^{\delta}> + ~ \dots ~ + <E|E^{\delta}>|^2 \simeq
(\cos \delta_x P/2\hbar + \cos \delta_p L/2\hbar)^2/4$.
To get this simple result we have ignored both the additive corrections
(such as $<N|E^{\delta}>$) that are small in the sparse limit, and
a multiplicative correction $\sim |<E|E^{\delta}>|^2$ that is very
close to unity in the limit of small displacements
$|\vec \delta| \ll \sqrt \hbar$.
Striking kinship of this form of the overlap with the interference term,
Eqs. (10) and (11), is no accident: The magnitude of the shifts that produce
orthogonality are $\delta_x =2 \pi \hbar /P$, $\delta_p = 2\pi \hbar 
/L$. Thus, displacing the states by the size of the basic
`tile' in the central interference pattern defined by Eqs. (11) and (12)
suffices to cause decoherence, if the environment starts in the $NWSE$ state.
larger shifts.}
\end{figure}

\newpage

\begin{figure}[h]
\centering
\epsfig{file=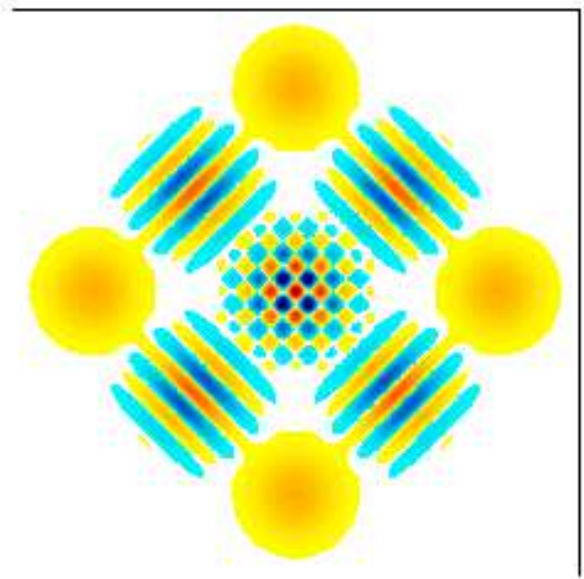, width=10cm, clip=}
\caption{
Snapshots of area $2 \pi \hbar$ extracted from Fig. 1a-c.
Smalest scales saturate on sub-Planck $a$, so that -- as can be estimated from
Fig. 1 -- the individual patches in the figure above appear on the scale
consistent with Fig. 1, 2, and Eq. (3). Such small phase space substructure
is physically significant: Displacement of the order of $\sqrt a$ -- size of
a typical patch -- suffices to make the perturbed state approximately
orthogonal to its old (unperturbed) self. Thus, $a$ sets the limit on
the sensitivity of the state to perturbations, and is therefore relevant for
decoherence. Note that the overlap of the original and displaced states
will behave differently when the interference pattern is irregular (as it is
here) rather than essentially periodic (as was the case in the compass state of
Fig. 2). This can be understood by expressing the state as a superposition
$|\psi> = \sum_k^N \alpha_k |k>$ of identical minimum uncertainty Gaussians
$|k> = G(x-x_k, p-p_k)$, each centered on $x_k,~p_k$ (see Eq. (6)).
In the sparse limit (i.e., when the Gaussians in the superposition
of $|\psi>$ do not overlap significantly) the overlap of the original and
displaced states is approximately:$|<\psi|\psi^{\delta}>| \simeq |\sum_k^N |\alpha_k|^2
e^{i(\delta_p x_k + \delta_x p_k) /\hbar}| =
|\sum_k^N w_k e^{i\phi_k(\vec \delta)}|$,
where $w_k = |\alpha_k|^2, \ \sum_k^Nw_k=1$ are the weights -- probabilities --
of finding the system in different states $|k>$. It is obvious that, when many
sparsely distributed Gaussians participate in $|\psi>$, so that $w_k \sim 1/N$,
size of the overlap is given by the distance covered by an eventually random
walk in a complex plane where individual steps have magnitude $w_k$ and
directions determined by phases $\phi_k$. Hence, as $\phi_k$ become random,
the overlap will rapidly decrease from unity to approximately $1/\sqrt N$,
and will likely remain small. (Such sums are a standard way of recovering
a Gaussian probability distribution, and have already been
studied in the context of decoherence some time ago$^4$.)
environment,
}\end{figure}

\end{document}